\pdfoutput=1

\documentclass[12pt]{article}
\usepackage{amsmath,epsf,amssymb,latexsym,cite,shadow,eucal,theorem,enumerate}
\usepackage[matrix,arrow,frame,import,curve,color]{xy}

\usepackage{graphicx}

\newtheorem{theorem}{Theorem}

\newtheorem{definition}{Definition}
{\theorembodyfont{\rmfamily}}

\newif\iffigs\figstrue

\DeclareFontFamily{U}{rsf}{}
\DeclareFontShape{U}{rsf}{m}{n}{
  <5> <6> rsfs5 <7> <8> <9> rsfs7 <10-> rsfs10}{}
\DeclareMathAlphabet\Scr{U}{rsf}{m}{n}

\def\O{\Scr{O}}
\def\C{{\mathbb C}}
\def\P{{\mathbb P}}

\def\Z{{\mathbb Z}}

\def\Im{\operatorname{Im}}
\def\im{\operatorname{im}}

\def\Hom{\operatorname{Hom}}

\def\Ext{\operatorname{Ext}}

\def\End{\operatorname{End}}

\def\coker{\operatorname{coker}}

\def\Spec{\operatorname{Spec}}

\def\Gr{\operatorname{Gr}}

\def\Sl{\operatorname{SL}}

\def\GU{\operatorname{U{}}}

\def\gd{\operatorname{gldim}}

\def\LT{\operatorname{\mathrm{LT}}}

\def\id{{\mathbf{1}}}

\def\CY{Calabi--Yau}
\def\LG{Landau--Ginzburg}

\def\cM{{\Scr M}}

\def\cL{{\Scr L}}
\def\cE{{\Scr E}}
\def\cF{{\Scr F}}

\def\cW{{\mathcal{W}}}
\def\DC{\mathbf{D}}

\def\QED{$\quad\blacksquare$}
\def\ff#1#2{{\textstyle\frac{#1}{#2}}}

\def\mf#1{\mathfrak{#1}}

\def\eqn#1#2{\begin{equation}#2
  \ifx{#1}{}\else\label{#1}\fi\end{equation}}

\begin{document}

\begin{titlepage}
\begin{flushright}
May 2010
\end{flushright}
\vspace{.5cm}
\begin{center}
\baselineskip=16pt
{\fontfamily{ptm}\selectfont\bfseries\huge
Quivers from Matrix Factorizations\\[20mm]}
{\bf\large  Paul S.~Aspinwall and David R.~Morrison
 } \\[7mm]

{\small

Center for Geometry and Theoretical Physics, 
  Box 90318 \\ Duke University, 
 Durham, NC 27708-0318 \\ \vspace{10mm}

Departments of Mathematics and Physics,\\
University of California,\\
Santa Barbara, CA 93106\\ \vspace{6pt}

 }

\end{center}

\begin{center}
{\bf Abstract}
\end{center}
We discuss how matrix factorizations offer a practical method of
computing the quiver and associated superpotential for a hypersurface
singularity. This method also yields explicit geometrical
interpretations of D-branes (i.e., quiver representations) on a
resolution given in terms of Grassmannians. As an example we analyze
some non-toric singularities which are resolved by a single $\P^1$ but
have ``length'' greater than one.  These examples have a much richer
structure than conifolds. A picture is proposed that relates matrix
factorizations in \LG\ theories to the way that matrix factorizations
are used in this paper to perform noncommutative resolutions.


\end{titlepage}

\vfil\break


\section{Introduction}    \label{s:intro}

To a Gorenstein singularity in three complex dimensions can be associated
 a quiver together with a superpotential in two equivalent ways.  To
simplify discussion let us assume this singularity is isolated.
First, using string theory language one may consider a stack of
D3-branes at the singular point. Then, as discussed in
\cite{kw,nonspherI}, one may associate a 4-dimensional field theory
to the near-horizon geometry using the AdS-CFT correspondence. This
four dimensional field theory is an $N=1$ gauge theory. The gauge
group and matter content is described by a quiver and the field theory
comes with a superpotential.

Alternatively, using more purely mathematical ideas, suppose the
singularity admits a crepant resolution $X$. Then the following picture
appears to be generally true.  The bounded derived category $\DC(X)$
of $X$ is equivalent to the bounded derived category of
representations of some quiver $Q$ with relations. The relations are
given by the derivatives of a single polynomial --- the
superpotential.

The equivalence of these two pictures arises from the derived category
description of B-type D-branes as proposed in \cite{douglas} (for a
review, see \cite{Aspinwall:2004jr}). The superpotential arises from the
$A_\infty$ structure of the derived category as discussed, for
example, in \cite{AF:superq,MR2450725,ginzburg06:CYa}.

It should be noted that the quiver and superpotential are not unique
for a given singularity. It was noted in \cite{BD:tilt} that this
ambiguity is a form of Seiberg duality \cite{Seiberg:1994pq}. It arises
from the fact that two quivers (and sets of relations) can yield
equivalent derived categories.

The question of interest in this paper is how does one compute a
quiver together with its superpotential from a given
singularity. There has been considerable work devoted to this question in the
past several years. A small sample of such work is given in
\cite{BGLP:dPo,Aspinwall:2004bs,FHH:toric,Hanany:2006nm,Aspinwall:2008jk,AgB:SuperP}.

A purpose here is to give a direct, mathematically rigorous method
that should work for {\em any\/} hypersurface singularity of dimension 3. It
should also be straight-forward to extend this method to complete
intersections. The restriction of the dimension to 3 arises
because it is only in this dimension that a superpotential can be
associated to the quiver relations.

The general idea comes from Van den Bergh's notion of a noncommutative
crepant resolution \cite{MR2057015,bergh04:nc}. This, in turn, arises
from notions in D-brane physics as observed by Berenstein and Leigh
\cite{Berenstein:2001jr}. 

Noncommutative resolutions are related to maximal Cohen--Macaulay
(MCM) modules, which, in turn, are related to matrix factorizations
via Eisenbud's work \cite{MR570778}. Associating matrix factorizations to
hypersurface singularities is not new --- it was described in
\cite{CMI,yoshino} for example. The purpose of this paper is to
see if it can be practically applied to yield a quiver and relations
for interesting examples.

The easiest example of a crepant resolution has a single $\P^1$ as the
exceptional set as analyzed in \cite{[L]}. While the humble conifold
is an example of such a singularity, there are many more richer
examples. The idea of ``length'' \cite{[Kol]} can be used as a partial
classification as we will explain in section \ref{ss:length}. Our examples will
concern cases of length 2.

The normal bundle of a $\P^1$ as the exceptional set is of the form
$\O(-1,-1)$, $\O(-2,0)$ or $\O(-3,1)$. The latter is by far the richest
and a classification of singularity types in the class is, as yet,
unknown although some progress has been made in this direction
\cite{gorenstein-weyl}. We will consider two sets of such
singularities and compute the quivers together with the
superpotentials. 

It is also of interest to understand the precise geometry of certain
basic D-branes. Of particular interest are the ``fractional branes''
into which a 0-brane may decay when it coincides with the
singularity. In quiver language these fractional branes are the
one-dimensional quiver representations. Specializing to one example we
will show how to explicitly find the objects in the derived category
associated to these D-branes. We will see how one of the fractional
branes corresponds to a sheaf that cannot be associated to a vector bundle.

Matrix factorizations are known to describe D-branes in the context of
\LG\ theories \cite{Kapustin:2003rc}. It is very interesting to
compare our use of matrix factorizations with those associated to \LG\
theories and we will discuss a possible relation later in this paper.


\section{Noncommutative Resolutions}  \label{s:ncr}

\subsection{The resolution}   \label{ss:res}

Let $\C^4$ have coordinates $(w,x,y,z)$ and let $S=\C[w,x,y,z]$. Let
$f\in S$ be a
polynomial such that $f=0$ has an isolated singularity at the
origin. Put
\begin{equation}
  R = \frac{S}{(f)},
\end{equation}
and $Y=\Spec R$ is a singular affine variety.

Let $M$ be a finitely generated $R$-module and let $\dim(M)$ be the
length of the shortest projective resolution of $M$.
The global dimension, $\gd R$, is defined as the largest value of
$\dim(M)$ for all such modules. 

Because $Y$ is not smooth, $\gd R$ is not finite. That is, there
are finitely-generated $R$-modules $M$ which have no finite projective
resolution. According to \cite{MR570778}, there is
a minimal free resolution of such
an $M$ which takes the form
\begin{equation}
\xymatrix@1{
\cdots\ar[r]&R^{\oplus d}\ar[r]^-\Psi&R^{\oplus d}\ar[r]^-\Phi&
R^{\oplus d}\ar[r]^-\Psi&R^{\oplus d}\ar[r]^-\Phi&\cdots\ar[r]&
R^{\oplus n_1}\ar[r]&M.}
\end{equation}
That is, the resolution becomes asymptotically periodic with period
2. Lifting the maps $\Psi$ and $\Phi$ to $S$ we have matrices with
polynomial entries. These matrices yield matrix factorizations:
\begin{equation}
  \Phi\Psi = \Psi\Phi = f.\id.  \label{eq:mf}
\end{equation}
Conversely, given a matrix factorization of the form (\ref{eq:mf}), we
may write $M=\coker \Psi$ to find a module which has an infinite free
resolution. 

Any module of the form $M=\coker \Psi$ coming from a matrix
factorization is a ``Maximal Cohen--Macaulay'' (MCM) module. An MCM
module is defined in terms of module depths but the only salient fact
for our purposes is that an MCM module always arises from
a matrix factorization,
where we include the trivial case $\Phi=1$, $\Psi=f$, yielding the
possibility $M\cong R$.

Let $M$ be an MCM-module (corresponding to a non-trivial matrix
factorization) and define
\begin{equation}
   A_1 = \End_R(R\oplus M).
\end{equation}
$A_1$ is a noncommutative $\C$-algebra and should be considered as a
non-commutative ``enhancement'' of the original coordinate ring
\begin{equation}
  R = A_0 = \End_R(R),
\end{equation}
by the MCM module $M$. $A_1$ may or may not have finite global
dimension. If it does not, one looks for further MCM modules to
produce
\begin{equation}
  A = A_d = \End_R(R\oplus M_1 \oplus M_2 \oplus \ldots\oplus M_d).
\end{equation}
Ultimately one hopes that $A$ has finite global dimension for large
enough $d$. When this happens one defines \cite{bergh04:nc} the
noncommutative resolution of $Y$ as $A$.

Let $\pi:X\to Y$ be any crepant resolution of $Y$. It is conjectured
that $\DC(X)$ is then equivalent to $\DC(\textrm{mod-$A$})$, the bounded
derived category of finitely generated right $A$-modules. It is in
this sense that $A$ represents a resolution of $Y$. This conjecture
has been proven in \cite{bergh04:nc} for the case of one-dimensional
exceptional sets in dimension 3, which is the case of interest in this
paper.

\subsection{Path algebras and superpotentials}  \label{ss:path}

Consider the $\C$-algebra
\begin{equation}
  A = \End_R(M_0\oplus M_1 \oplus M_2 \oplus\ldots\oplus M_m),
\end{equation}
where $M_0=R$. We compose homomorphisms right to left as usual. This
algebra may be viewed as the path algebra of a quiver $Q$ with $m$
vertices and some relations. Let $c_j\in A$ be the idempotent element
corresponding to the identity in $\Hom_R(M_j,M_j)$. This is a path of
length zero at node $j$. Define
\begin{equation}
  P_j = c_jA.
\end{equation}
This may be viewed as the space generated by all paths {\em ending\/}
at node $j$ and is a projective right $A$-module. If $\gamma$ is a
path from node $i$ to node $j$ then $\gamma P_i\subset P_j$ and this
path gives a homomorphism from node $i$ to node $j$. Thus the $M_j$'s
in the tilting module $M_0\oplus M_1 \oplus M_2 \oplus\ldots\oplus
M_m$ correspond to the $P_j$'s in the category of right
$A$-modules.\footnote{There are two conventions in the literature. One
  is to use right $A$-modules. If quiver representations are viewed as
  left $A$-modules then the direction of the arrows in $Q$ must be
  reversed. By using right $A$-modules we keep the matrix
  multiplication for the rest of the paper in the conventional direction!}

We are interested in relations of a specific type, namely those coming
from a {\em superpotential}. Let $A_{\textrm{cyc}}$ denote the quotient
of a subalgebra of $A$ generated by cycles in the quiver, where we
identify cyclic permutations of arrows. The superpotential is an element
\begin{equation}
  \cW \in A_{\textrm{cyc}},
\end{equation}
such that the relations of the quiver are generated by the cyclic
derivatives of $\cW$. That is, if arrows in the quiver are denoted
$a_i$, one defines
\begin{equation}
  \frac{\partial}{\partial a_j}
  a_{i_1}a_{i_2}\ldots a_{i_n} =
   \sum_{s,i_s=j} a_{i_{s+1}}a_{i_{s+2}}\ldots a_{i_n}
                 a_{i_1}\ldots a_{i_{s-1}},
\end{equation}
(modifying in the obvious way if $j=i_1$ or $i_n$) and extends this by
linearity to $\cW$.

Let us emphasize that we are defining the superpotential in terms of
the relations. There is another definition in terms of the
$A_\infty$-algebra of the derived category $\DC(X)$ which is more
directly associated to the physics definition. The equivalence of
these definitions has been proved in some cases
\cite{AF:superq,MR2450725}. Having determined the superpotential for a
given example in terms of relations one can then verify that it agrees
with the $A_\infty$-algebra as we discuss in section \ref{ss:Ainf}.

\def\AA{\hbox{$A$--$A$}}

$A$ is an $\AA$ bimodule. A projective $\AA$ bimodule is a summand of
$A\otimes_\C A$ and so $A$ itself is not, in general, a projective
$\AA$ bimodule. Suppose we have a projective resolution
\begin{equation}
\xymatrix@1{
  \ldots\ar[r]&B_2\ar[r]&B_1\ar[r]&B_0\ar[r]&A\ar[r]&0.}  \label{eq:AAres}
\end{equation}
Let $M$ be a right $A$-module. It was shown in 
\cite{MR1432347} that
\begin{equation}
\xymatrix@1{
  \ldots\ar[r]&M\otimes_AB_2\ar[r]&M\otimes_AB_1\ar[r]&
              M\otimes_AB_0\ar[r]&M\ar[r]&0,}
\end{equation}
is a resolution by projective right $A$-modules of $M$. Thus, if we
can find a resolution of length $d$ for (\ref{eq:AAres}) then the
global dimension of $A$ can be no greater than $d$.

In fact, \cite{MR1432347} gives a prescription for constructing the resolution
(\ref{eq:AAres}) up to $B_2$. In the special case that the relations
come from a superpotential one can write down a specific candidate for a
projective resolution \cite{Bergman:2006gv}. Introduce the projective bimodules
\begin{equation}
  P_{ij} = Ac_i\otimes c_jA.
\end{equation}
Now we have a resolution
\begin{equation}
\xymatrix{
  *!(0,1.8)\txt{$\displaystyle{\bigoplus_{v\in\Gamma_0} P_{vv}}$}
            \ar[r]^-{\phi_3}&
  *!(0,1.8)\txt{$\displaystyle{\bigoplus_{a\in\Gamma_1} P_{o(a),t(a)}}$}
            \ar[r]^-{\phi_2}&
  *!(0,1.8)\txt{$\displaystyle{\bigoplus_{a\in\Gamma_1} P_{t(a),o(a)}}$}
            \ar[r]^-{\phi_1}&
  *!(0,1.8)\txt{$\displaystyle{\bigoplus_{v\in\Gamma_0} P_{vv}}$}
            \ar[r]^-{m}&
  A.
}  \label{eq:Pres}
\end{equation}

The maps are defined as follows. $m$ is simply multiplication
$m:A\otimes A\to A$. We also have
\begin{equation}
  \phi_1(c_{t(a)}\otimes c_{o(a)}) = c_{t(a)}\otimes a -
          a \otimes c_{o(a)}.
\end{equation}
The relations are in one-to-one correspondence with the arrows since a
relation is obtained from $\partial \mathcal{W}/\partial a$. A
relation $r$ is a linear combination of paths starting at $t(a)$ and
ending at $o(a)$. If one such path is given by $a_ma_{m-1}\ldots a_1$
then
\begin{equation}
\begin{split}
  \phi_2(a_ma_{m-1}&\ldots a_1) =
    c_{t(a_m)} \otimes a_{m-1}\ldots a_1 \,+\\
    &\sum_{i=2}^{m-1} \Bigl(a_ma_{m-1}\ldots a_{i+1}\otimes a_{i-1}
          \ldots a_1\Bigr)
    + a_ma_{m-1}\ldots a_2\otimes c_{o(a_1)}.
\end{split}
\end{equation}
$\phi_2$ is then defined on relations by extending by linearity. Finally
\begin{equation}
  \phi_3(c_v\otimes c_v) = \sum_{t(a)=v}a\otimes c_{t(a)}
       - \sum_{o(a)=v}c_{o(a)}\otimes a,
\end{equation}
where the image of the summands on the right end up in the
relation associated to the arrow $a$.

Applying the functor $c_i\otimes_A-$ to (\ref{eq:Pres}) we obtain
\begin{equation}
\xymatrix{
  P_i
            \ar[r]^-{\psi_3}&
  *!(0,1.8)\txt{$\displaystyle{\bigoplus_{o(a)=i} P_{t(a)}}$}
            \ar[r]^-{\psi_2}&
  *!(0,1.8)\txt{$\displaystyle{\bigoplus_{t(a)=i} P_{o(a)}}$}
            \ar[r]^-{\psi_1}&
  P_i
            \ar[r]&
  L_i,
}  \label{eq:Lres}
\end{equation}
where $L_i=c_i$ is the ``vertex simple'' one-dimensional representation of 
$A$ associated to the vertex $i$. (\ref{eq:Lres}) is a projective
resolution of $L_i$ if (\ref{eq:Pres}) is exact. The maps $\psi_1$ and
$\psi_3$ consist of appending the corresponding arrows in the sum. The
map $\psi_2$ is a matrix whose $(i,j)$th entry is the part of the relation
$\partial\mathcal{W}/\partial a_i$ ending with $a_j$ with $a_j$ removed.

Whether (\ref{eq:Pres}) really is an exact sequence depends on the
superpotential $\cW$. It was shown in \cite{ginzburg06:CYa} that this
is equivalent to $A$ being a ``\CY\ algebra'' of dimension 3. That is,
$A$ admits something similar to a Serre functor modified for the fact
that $X$ is not compact. We refer to \cite{ginzburg06:CYa} for the
details of the definition. The Serre functor corresponds to the
existence of spectral flow in the associated conformal field theory and
so corresponds to the fact we have a supersymmetric solution.

An analysis of the exactness of (\ref{eq:Pres}) may be performed by using
noncommutative Gr\"obner basis methods as described in
\cite{MR1714602}, for example. Let $\mathsf{P}$ be the path algebra
before the relations are imposed and let $I$ be the double-sided ideal
of relations. One needs to find a Gr\"obner basis, $\mathcal{G}$, for
$I$. This can be attempted using methods described in \cite{MR1714602}
but there is no guarantee that the resulting 
basis will have a finite number of
elements. Indeed, for the examples in this paper $\mathcal{G}$ is not
finite. However, given a systematic presentation of $\mathcal{G}$ one
can proceed to analyze (\ref{eq:Pres}).

The first step in analyzing a Gr\"obner basis is a choice of monomial
ordering on $\mathsf{P}$. We use the length-right-lexicographic
ordering. First a choice of ordering on the vertices and on the arrows
of the quiver is made with all vertices less than all arrows.  Paths
are then ordered by length. Then, for paths of equal length we use
a lexicographic ordering reading right-to-left using the chosen order
on the arrows (and vertices for zero-length paths).

This monomial ordering on the algebra $\mathsf{P}$ can be extended to an
ordering on the $\mathsf{P}$-modules in the resolution
(\ref{eq:Pres}) following \cite{MR1432347}. Consider the
$\bigoplus_{a\in\Gamma_1} P_{t(a),o(a)}$ term first. Suppose
\begin{equation}
\begin{split}
   q\otimes p &\in P_{t(a_i),o(a_i)}\\
   s\otimes r &\in P_{t(a_j),o(a_j)}.
\end{split}
\end{equation}
We then impose an ordering by the following sequence of rules
\begin{enumerate}
\itemsep=0mm
\item $l(q) <  l(s)$
\item $l(p)>l(r)$
\item $q<s$
\item $p>r$
\item $a_i>a_j$.
\end{enumerate}
In each, the subsequent rule is applied if the preceding inequality
test results in equality. The same ordering is applied for
$\bigoplus_{a\in\Gamma_0} P_{vv}$ except the last test is irrelevant.
The key property required is that
\begin{equation}
  \LT(x) > \LT(y) \;\;\Rightarrow\;\; \LT(\phi_n(x)) > \LT(\phi_n(y)),
  \qquad\forall n.  \label{eq:order}
\end{equation}
This is immediately true for $\phi_3$ and $\phi_1$. By choosing an
appropriate modification of rule 5 for $\bigoplus_{a\in\Gamma_1}
P_{o(a),t(a)}$ depending on the details of the relations, it can be
made true for $\phi_2$.

\begin{theorem}
  The complex (\ref{eq:Pres}) is exact if and only if the complexes
associated to the vertex simples (\ref{eq:Lres}) are exact for all
nodes in the quiver.
\end{theorem}

This is proved as follows. We already know that (\ref{eq:Lres}) is
exact if (\ref{eq:Pres}) is exact from the exactness of $M\otimes_A-$.
To prove that (\ref{eq:Pres}) is exact we need to show that
$\ker(\phi_2)\subset\im(\phi_3)$ and
$\ker(\phi_1)\subset\im(\phi_2)$. Suppose $x\in\ker(\phi_2)$. From
(\ref{eq:order}) we know that $\LT(\phi_2(x))$ cannot cancel with any
other term in the image of $x$. If $\LT(x)=q\otimes p$ then we know from
the ordering that $\LT(\phi_2)$ is of the form $q\otimes zp$ for some
path $z$ arising from the relations. But this is exactly the part of
the map $\phi_2$ which is seen by the resolutions of the vertex
simples (\ref{eq:Lres}). That is, if all the resolutions (\ref{eq:Lres}) 
are exact it must be that $zp\in I$ implies that $p$ may be reduced
(with respect to the ordering) modulo terms in the image of $\psi_3$.
Thus $x$ may be reduced modulo the image of $\phi_3$. Thus, by
induction, $x$ is in the image of $\phi_3$. Thus
$\ker(\phi_2)\subset\im(\phi_3)$. The same argument shows
$\ker(\phi_1)\subset\im(\phi_2)$.  \QED

This yields an algorithm for computing the noncommutative resolution
of a hypersurface singularity:
\begin{enumerate}
\item Begin with $T=R$.
\item Compute the quiver and relations for the path algebra
  $A=\End_R(T)$.
\item If the relations which arise are compatible with a superpotential,
  check if the complexes (\ref{eq:Lres}) are all exact. If they are, we stop.
\item The global dimension of $A$ is presumably infinite so we can
  find an irreducible module $M$ with no finite projective resolution.
\item Replace $T$ by $T\oplus M$ and repeat from step 2.
\end{enumerate}
Unfortunately, we don't have a proof that this algorithm terminates in
general.

\subsection{Maps between cokernels}  \label{ss:maps}

One can be quite systematic about computing the quiver and looking for
relations. Suppose we have two $R$-modules of the form
$M=\coker \Psi$ and $M'=\coker \Psi'$ where $\Psi$ is an $n\times n$
matrix and $\Psi'$ is an $n'\times n'$ matrix.
Any homomorphism from $M$ to $M'$ lifts to a homomorphism
$\alpha:R^n\to R^{n'}$.  The converse is trickier, however:  a
given homomorphism
$\alpha:R^n\to R^{n'}$ descends to a homomorphism $M\to M'$
if and only if there exists $\alpha':R^n\to R^{n'}$ such that
\[ \alpha  \Psi = \Psi'  \alpha'. \]
(This guarantees that the image of $\Psi$ is mapped to the image of
$\Psi'$ by $\alpha$.)  Note that $\alpha'$ won't in general be unique.

Our main concern in this paper will be with a {\em single\/} matrix
factorization of the form (\ref{eq:mf}).  We build a natural quiver
with two nodes from this matrix factorization, with one node
representing $M=\coker \Psi$ and the other representing the rank $1$
$R$-module $R$ (which can be written as $\coker {(0)}$).

Homomorphisms from $R\to M$ are given by arbitrary $n\times1$ matrices,
and we can use as a generating set for these the standard basis (column)
vectors for $R^n$: namely, $e_1$, $e_2$, \dots, $e_n$.
However, if a homomorphism's image is contained in the image of $\Psi$
(i.e., in the column space of the matrix),
then the homomorphism is trivial

Homomorphisms from $M\to R$ are given by $1\times n$ matrices $\alpha$
such that $\alpha\Psi=0$; a basis for those is given by the rows
of $\Phi$, which we can represent in the form $\alpha_i = e_i^T\Phi$.

In most cases, there will be additional endomorphisms of $M$ or of $R$
which need to be added to this quiver (and which may allow us to
eliminate some of the $\alpha_i$ and $e_j$), but for the moment, we
consider the algebra $\mathcal B$ generated by the maps $\alpha_i$ and
$e_j$.  The task is to find the abstract relations which
these maps satisfy, and to determine whether the resulting algebra is
the full endomorphism algebra of $R\oplus M$ or not.


There is one type of relation in the algebra $\mathcal B$ which
is always present: these are relations derived from the fact that endomorphisms
of $R$ are given by $1\times1$ matrices, so they must commute.
That is,
\begin{equation}\label{eq:rel1}
 \alpha_i e_j \alpha_k e_\ell = \alpha_k e_\ell \alpha_i e_j
\end{equation}
for all $i$, $j$, $k$, $\ell$.
Note that $\alpha_i e_j$ is represented by the $1\times1$ matrix
$[\Phi_{ij}]$.

The next three types of relations are derived from two specific properties
which the given matrix factorization may have.  First, if there are
linear relations among the matrix entries $\Phi_{ij}$, then the
corresponding relations among the $\alpha_ie_j$'s must hold in
the algebra $\mathcal B$.  Second, if every entry in $\Psi$ is a
linear combination of entries of $\Phi$, then we get a relation from
each row of $\Psi$ by the following construction.
Since $e_i^T\Psi = \sum_j \Psi_{ij}e_j^T$, we see 
\[0=e_i^T\Psi\Phi=              
\sum \Psi_{ij}e_j^T\Phi = \sum_j \Psi_{ij}\alpha_j.\]
Writing 
\[\Psi_{ij}=\sum_{k,\ell} c_{ij}^{k\ell}\Phi_{k\ell}=\sum_{k,\ell} c_{ij}^{k\ell}\alpha_ke_\ell,\]
we find a relation
\[0 = \sum_{j,k,\ell} c_{ij}^{k\ell}\alpha_ke_\ell\alpha_j\]
for each $i$.
And third, again when every entry in $\Psi$ is a linear combination
of entries of $\Phi$, we get a relation from each column of $\Psi$
as follows.  We have $\Psi e_j = \sum_i e_i\Psi_{ij}$ and this 
combination is the zero homomorphism $R\to M$ since its image lies
in the column space of $\Psi$.  Thus,
\[0 = \sum_i e_i\Psi_{ij} = \sum_{i,k,\ell}e_i(c_{ij}^{k\ell}\Phi_{k\ell})
= \sum_{i,k,\ell} c_{ij}^{k\ell} e_i\alpha_ke_\ell.\]


\section{Easy Examples}  \label{s:len}

\subsection{Length} \label{ss:length}

Let $M$ be an $R$-module. Let us assume both $R$ and $M$ are
Noetherian. A {\em composition series\/} of length $n$ for $M$ is a
strictly decreasing chain of submodules
\begin{equation}
  M = M_0 \supset M_1 \supset \ldots \supset M_n=0,
\end{equation}
such that $M_j/M_{j+1}$ is nonzero and has no nonzero proper
submodules for all $j$.

It can be shown that all compositions series for a fixed $M$ have the
same length (see, for example, \cite{Eis:CA}) and this defines the
{\em length of $M$.} Similarly one can define the length of coherent sheaves.

Now consider a resolution of singularities
\begin{equation}
\pi: X \to Y,
\end{equation}
where $Y=\Spec R$ has an isolated singularity at $p$. Let $\mf{m}$ denote
the maximal ideal in $R$ associated to the point $p$. The inclusion
map $\mf{m}\to R$ pulls back via $\pi^*$ to a map of
sheaves\footnote{If $M$ is an $R$-module we will also use $M$ to
  denote the associated coherent sheaf on $\Spec R$.}
\begin{equation}
  f:\pi^*\mf{m} \to \O_X.
\end{equation}
Let $\cE$ denote the sheaf on $X$ associated to $\coker f$. $E$, the
support of $\cE$, is then the exceptional set of the
resolution. $E$ decomposes into a union of irreducible components:
\begin{equation}
  E = \bigcup_i E_i.
\end{equation}
Correspondingly we have a primary decomposition of the module
associated to $\cE$:
\begin{equation}
  \cE = \bigcap_k M_k.
\end{equation}
Let $M_i$ be an element of the above primary decomposition
corresponding to a minimal associated prime of $\cE$. Then $M_i$ is associated
to one of the components $E_i\subset E$. We then have
\begin{definition}
  The {\em length\/} of an irreducible component $E_i\subset E$ of the
  exceptional set is defined as the length of the module $M_i$.
\end{definition}
One may regard the length as the multiplicity of a component of the
exceptional set for a resolution. 

In this paper our interest concerns resolutions with a single $\P^1$
as the exceptional set. Such resolutions are therefore labeled by a
single length.

Cases where lengths of greater than one are very common but generally
harder to analyze. For example, suppose that the resolution
$\pi:X\to Y$ is described torically. That is, it is obtained by some
subdivision of a fan $\Sigma$. The explicit form of such a resolution
in terms of coordinate patches (see, for example, \cite{Fulton:})
immediately implies that only components of length one can ever appear
in the exceptional set. Similarly any generic deformation of a toric
case will still only exhibit length one. Thus, the various toric
methods that can apply for computing quivers, etc., cannot be used in
any direct way to study lengths greater than one. 

\subsection{Du Val singularities}  \label{ss:dV}

A useful and straight-forward example of higher lengths concerns the
well-known ADE resolutions of $\C^2/G$, for $G\subset\Sl(2,\Z)$ as
studied in in the McKay correspondence in
\cite{McKay:,MR740077,math.AG/9812016}. In particular, for the current
context, see \cite{Aus:McKay,AR:McKay}. Here, the lengths are the Kac
labels on the Dynkin diagram. E.g., the resolution associated to $E_6$
has lengths
\begin{equation}
\begin{xy} <1.3mm,0mm>:
  (0,0)*++={\scriptstyle 1}*\frm{o}="a",
  (10,0)*++={\scriptstyle 2}*\frm{o}="b",
  (20,0)*++={\scriptstyle 3}*\frm{o}="c",
  (30,0)*++={\scriptstyle 2}*\frm{o}="d",
  (40,0)*++={\scriptstyle 1}*\frm{o}="e",
  (20,10)*++={\scriptstyle 2}*\frm{o}="f"
  \ar@{-}"a";"b"
  \ar@{-}"b";"c"
  \ar@{-}"c";"d"
  \ar@{-}"d";"e"
  \ar@{-}"c";"f"
\end{xy}.  \label{eq:E6}
\end{equation}

The description of these Du Val singularities in terms of
noncommutative resolutions and matrix factorizations is implicit in
\cite{MR740077} and described in \cite{flopfactorization}.

Let $G$ act on $\C^2=\Spec(\C[s,t])$ and thus $\C[s,t]$. $Y=\C^2/G$
then has coordinate ring $R=\C[s,t]^G$.  Denote by $\rho_i$ the
nontrivial irreducible representations of $G$. There is then a
decomposition
\begin{equation}
  \C[x,y] = R \oplus\left(\bigoplus_i M_i\otimes\rho_i\right),
\end{equation}
where $M_i$ are $R$-modules. All such $M_i$ can be associated to matrix
factorizations as listed in \cite{MR740077,math.AG/9812016},
\begin{equation}
  M_i = \coker \Psi_i,
\end{equation}
where $\Psi_i$ is a $2l_i\times 2l_i$ matrix with entries in $R$. It
turns out, from the classification in \cite{MR740077}, that each
$M_i$ is associated to an irreducible component of the exceptional
divisor with length $l_i$.

A noncommutative resolution of $\C^2/G$ is then described by the path
algebra of the McKay quiver which is also given by
\begin{equation}
  A = \End_R\left(R\oplus\left(\bigoplus_i M_i\right)\right).
\end{equation}

\subsection{A Flop}  \label{ss:flop1}

Consider the hypersurface singularity
\begin{equation}
  x^2 - y^{2k} + wz = 0,
\end{equation}
where $k$ is a positive integer.
This admits a $2\times 2$ matrix factorization
\begin{equation}
  \Psi = \begin{pmatrix}w&-x-y^{k}\\x-y^{k}&z\end{pmatrix},\qquad
  \Phi = \begin{pmatrix}z&x+y^k\\-x+y^k&w\end{pmatrix}.
\end{equation}
Set $M=\coker\Psi$. The quiver for $A=\End_R(R\oplus M)$
is constructed as follows.

First note that a given morphism $f\in\Hom_R(M_1,M_2)$ can be multiplied
by any $r\in R$ giving $\Hom_R(M_1,M_2)$ the structure of an
$R$-module. For the cases considered this module is finitely-generated
and can be computed straight-forwardly using packages such as Macaulay
2. One obtains
\begin{equation}
  \Hom_R(R,R)\cong\Hom_R(M,M)\cong R,
\end{equation}
and both $\Hom(R,M)$ and $\Hom(M,R)$ are $R$-modules with two
generators. These generators are precisely the $e_i$'s and
$\alpha_j$'s of section \ref{ss:maps}. Thus we account for all paths
if we include the $e_i$'s, $\alpha_j$'s and all maps corresponding to
multiplication by $x$, $y$, $z$ and $w$.

It is easy to see that
\begin{equation}
\begin{split}
  \alpha_1e_1 &= z\\
  \alpha_1e_2 &= x+y^k\\
  \alpha_2e_1 &= -x+y^k\\
  \alpha_2e_2 &= w,
\end{split}
\end{equation}
and so all maps $R\to R$ are generated except for multiplication by
$y$ in the case $k>1$. We also have
\begin{equation}
  e_1\alpha_1 = \begin{pmatrix}z&x+y^k\\0&0\end{pmatrix}
   = \begin{pmatrix}z&0\\0&z\end{pmatrix}\pmod{\Im\Psi}
\end{equation}
and similarly maps from $M\to M$ corresponding to multiplication by
$w$, $x+y^k$ and $-x+y^k$ can be constructed. Thus, if $k>1$ we need
to add two arrows to the quiver to account for multiplication by
$y$. The resulting quiver is
\vspace{6mm}
\begin{equation}
\xymatrix@C=25mm{
  R\ar@/^/[r]|{\,e_2\,}\ar@/^7mm/[r]|{\,e_1\,}
          \ar@`{c+(-20,20),c-(0,40)}[]|{\vphantom{E_0^0}y_1}
&M\ar@/^/[l]|{\,\alpha_1\,}\ar@/^7mm/[l]|{\,\alpha_2}
          \ar@`{c+(20,20),c-(0,40)}[]|{\vphantom{E_0^0}y_2}
}\end{equation}
\vspace{5mm}

If $k=1$ these two extra loops are not added to the quiver.
There are now some obvious relations. Assuming $k>1$ we have
\begin{equation}
\begin{split}
  \alpha_1e_2 + \alpha_2e_1 &= 2y_1^k\\
  e_2\alpha_1 + e_1\alpha_2 &= 2y_2^k\\
  e_i y_1 &= y_2 e_i\\
  y_1 \alpha_i &= \alpha_i y_2.
\end{split}
\end{equation}
One can check these relations imply all the relations discussed in
section \ref{ss:maps}.  Furthermore, it is clear these relations can
be integrated up to form a superpotential
\begin{equation}
  \cW = y_1\alpha_1e_2 + y_1\alpha_2e_1 - \alpha_1y_2e_2
        - \alpha_2y_2e_1 - \ff2{k+1}y_1^{k+1} +
        \ff2{k+1}y_2^{k+1}.
\end{equation}
This recovers the known result of \cite{CKV:quiv,Aspinwall:2004bs}. Similarly
the case $k=1$ recovers the simple conifold quiver and relations of
\cite{kw,nonspherI}. 

\subsection{A generalized conifold}

Consider the hypersurface singularity
\begin{equation}
  f= x^3 - xy^2 -wz = 0.  \label{eq:gcon3}
\end{equation}
Let 
\begin{equation}
  M_1 = \coker \begin{pmatrix} x&-z\\-w&x^2-y^2 \end{pmatrix},
\end{equation}
corresponding to a matrix factorization of $f$. Now try the path
algebra $A=\End(R\oplus M_1)$. This yields a quiver
\vspace{2mm}
\begin{equation}
\xymatrix@C=30mm{
  R\ar@/^4mm/[r]|{\left(\begin{smallmatrix}1\\0\end{smallmatrix}\right)\,}
   \ar@/^10mm/[r]|{\left(\begin{smallmatrix}0\\1\end{smallmatrix}\right)\,}
          \ar@`{c+(-20,20),c-(0,40)}[]|{\vphantom{E_0^0}y}
&M_1\ar@/^4mm/[l]|{\,\left(\begin{smallmatrix}w&x\end{smallmatrix}\right)\,}
  \ar@/^10mm/[l]|{\,\left(\begin{smallmatrix}x^2-y^2&z\end{smallmatrix}\right)\,}
          \ar@`{c+(20,20),c-(0,40)}[]|
   {\left(\begin{smallmatrix}y&0\\0&y\end{smallmatrix}\right)}
}  \label{eq:q-part}
\end{equation}
\vspace{1mm}

Some relations correspond to the fact that $y$ commutes with paths of
length 2 going from $R$ to $M_1$ and back again. It is easy to see that
it is impossible for these relations to come from the derivatives of
any superpotential. Correspondingly, the global dimension of $A$ is
infinite.

The quiver (\ref{eq:q-part}) is therefore a {\em partial\/} noncommutative
resolution of the singularity (\ref{eq:gcon3}). There is another
matrix factorization yielding
\begin{equation}
  M_2 = \coker \begin{pmatrix} x^2+xy&-z\\-w&x-y \end{pmatrix}.
\end{equation}
Now set $A=\End(R\oplus M_1\oplus M_2)$. Then we obtain a quiver
\vspace{2mm}
\def\pvv{\vphantom{W_{1_1}}}
\begin{equation}
\xymatrix@C=20mm@R=20mm{
&R\ar@/^/[dl]|{\pvv\gamma_2}\ar@/^/[dr]|{\pvv\alpha_1}\\
M_2\ar@/^/[rr]|{\beta_2}\ar@/^/[ur]|{\pvv\gamma_1}&&
M_1\ar@/^/[ll]|{\beta_1}\ar@/^/[ul]|{\pvv\alpha_2}
}\end{equation}
where
\begin{equation}
\begin{alignedat}{2}
  \alpha_1 &= \begin{pmatrix}0\\1\end{pmatrix}&
  \alpha_2 &= \begin{pmatrix}w&x\end{pmatrix}\\
  \beta_1 &= \begin{pmatrix}x+y&0\\0&1\end{pmatrix}&\qquad
  \beta_2 &= \begin{pmatrix}1&0\\0&x+y\end{pmatrix}\\
  \gamma_1 &= \begin{pmatrix}x-y&z\end{pmatrix}&
  \gamma_2 &= \begin{pmatrix}1\\0\end{pmatrix}
\end{alignedat}
\end{equation}
There are obvious relations such as $\gamma_1\gamma_2\alpha_1\alpha_2=
\alpha_1\alpha_2\gamma_1\gamma_2$ but these cannot arise directly from
a superpotential since there is no loop at $R$. So we should look for
relations between paths of length 3. Writing down all such paths gives
relations arising from the superpotential
\begin{equation}
  \mathcal{W} = \alpha_1\alpha_2\alpha_1\alpha_2
+\beta_1\beta_2\beta_1\beta_2
+\gamma_1\gamma_2\gamma_1\gamma_2
+\alpha_2\alpha_1\gamma_1\gamma_2
+\beta_2\beta_1\alpha_1\alpha_2
+\gamma_2\gamma_1\beta_1\beta_2,
\end{equation}
in agreement with \cite{Gubser:1998ia}. This is the complete
resolution.


\section{Flops of Length Two}  \label{s:f2}

\subsection{The universal flop of length two} \label{ss:univ}

A threefold singularity $x\in X$ which has a crepant resolution $\pi:Y\to X$
such that $\pi^{-1}(x)$ is a rational curve of length two is always
part of a {\em flop} \cite{[Kol]}: there is a second crepant resolution
$\pi:Y^+\to X$ of the same singularity.

A ``universal flop of length two'' $\mathcal X$ was constructed in
\cite{flopfactorization}.  This is a hypersurface in $\mathbb{C}^7$
which has two small resolutions $\mathcal Y\to \mathcal X$ and 
$\mathcal Y^+\to\mathcal X$ determined from a matrix factorization
by a ``Grassmann blowup.''  This hypersurface is universal in the
sense that for every
threefold singularity $X$ which has a crepant resolution $Y\to X$
with exceptional set a rational curve of length two, there is a map
from $X$ to $\mathcal X$ such that $Y\to X$ is the pullback of 
$\mathcal Y\to\mathcal X$.  In particular,
all of the algebraic properties of the hypersurface in $\mathbb{C}^7$
will be inherited by any such threefold via the map $X\to\mathcal X$.

We now give a partial analysis of $\End(R\oplus M)$ for the
module $M$ obtained from the matrix factorization associated
to the universal flop of length two.  The generators and relations we
find will be a common feature of {\em any}\/ flop of length two.
In later sections we
will specialize to specific instances of the universal flop,
 which produce specific
flops, and we will refine the generators and relations of the
algebra in those cases.

  The  hypersurface equation in $\C^7$ which describes the universal
flop of length two is
\begin{equation}
f = x^2+uy^2+2vyz+wz^2+(uw-v^2)t^2. \label{eq:univ2}
\end{equation}
Put
$\Psi = xI+\Xi$
and $\Phi = xI-\Xi$ where
\begin{equation*} 
\Xi =
\begin{bmatrix}
-vt & y & z & t \\
-uy - 2v z & v t & -ut & z \\
-wz & wt & -vt & -y \\
-uwt & -wz & uy+2v z& v t \\
\end{bmatrix}.
\end{equation*}

The entries in $\Phi$ all occur as the entries in $1\times1$ matrices in
our algebra, since $\alpha_i e_j = e_i^T \Phi e_j = [\Phi_{ij}]$ 
(with $i$ and $j$ {\em fixed}).  There are linear relations among these:
essentially, some duplicates among the entries, up to sign.
When duplicate entries occur, we get a relation among our operators
$\alpha_i$ and $e_j$.
This gives the following relations:
\begin{equation}
\label{eq:rel2}
\begin{aligned}
\alpha_1 e_1 &= \alpha_3 e_3 
&\quad 
\alpha_1 e_2 &= -\alpha_3 e_4 
&\quad 
\alpha_1 e_3 &= \alpha_2 e_4
\\
\alpha_2 e_1 &= -\alpha_4 e_3 
&\quad
\alpha_2 e_2 &= \alpha_4 e_4
&\quad
\alpha_3 e_1 &= \alpha_4 e_2
\\
\end{aligned}
\end{equation}

There are four relations which we get from the columns of $\Psi$:
$\sum_j \Psi_{ij}\alpha_j=0$.  Using the fact that each $\Psi_{ij}$
coincides with some $\Phi_{k\ell}=\alpha_k e_{\ell}$, we can write
these relations as
\begin{equation}
\label{eq:rel3}
\begin{aligned}
0&=\alpha_4e_4\alpha_1+\alpha_3e_4\alpha_2-\alpha_2e_4\alpha_3-\alpha_1e_4\alpha_4\\
0&=\alpha_4e_3\alpha_1+\alpha_3e_3\alpha_2-\alpha_2e_3\alpha_3-\alpha_2e_4\alpha_4\\
0&=-\alpha_4e_2\alpha_1-\alpha_3e_2\alpha_2+\alpha_4e_4\alpha_3-\alpha_3e_4\alpha_4\\
0&=-\alpha_4e_1\alpha_1-\alpha_4e_2\alpha_2-\alpha_4e_3\alpha_3+\alpha_3e_3\alpha_4\\
\end{aligned}
\end{equation}
where we used the relations in eq.~\eqref{eq:rel2} to simplify these slightly.

There are also relations from the columns of $\Psi$, taking the form
$\sum_i e_i \Psi_{ij}=0$.  Again using the fact that each $\Psi_{ij}$
coincides with some $\Phi_{k\ell}=\alpha_k e_{\ell}$, we can write
these relations as
\begin{equation}
\label{eq:rel4}
\begin{aligned}
0&= e_1\alpha_4e_4+e_2\alpha_4e_3-e_3\alpha_4e_2-e_4\alpha_4e_1\\
0&= e_1\alpha_3e_4+e_2\alpha_3e_3-e_3\alpha_3e_2-e_4\alpha_4e_2\\
0&=-e_1\alpha_2e_4-e_2\alpha_2e_3+e_3\alpha_4e_4-e_4\alpha_4e_3\\
0&=-e_1\alpha_1e_4-e_2\alpha_2e_4-e_3\alpha_3e_4+e_4\alpha_3e_3\\
\end{aligned}
\end{equation}
where again we used the relations in eq.~\eqref{eq:rel2} to simplify.

There are at least two more endomorphisms of $M$ which so far have not been
accounted for.\footnote{We found these by making detailed calculations
in the full deformation of the D4 quiver which had been determined
in \cite{flopfactorization}.} We denote these
\begin{align*}
a = a' &=
\begin{bmatrix}
0&1&0&0\\-u&0&0&0\\-2v&0&0&1\\0&2v&-u&0\\
\end{bmatrix}
\\
b = b' &=
\begin{bmatrix}
0&0&1&0\\0&0&0&-1\\-w&0&0&0\\0&w&0&0\\
\end{bmatrix}
\\
\end{align*}
(Recall that for an endomorphism of $M$, lifted to a map
$E:R^4\to R^4$, we needed a second map $E'$ such that
$E\Psi = \Psi E'$.)  Note that for both of the above endomorphisms, since
$\Phi=2xI-\Psi$, we also have $\Phi E=E\Phi$.

These endomorphisms have two important properties (which easily follow
from $\Phi E = E\Phi$). First, letting
\[ c = \begin{bmatrix} x+vt & -y&-z&-t\\ \end{bmatrix} \]
be the first row of $\Phi$,
the rows of
$\Phi$ are $\alpha_1=c$, $\alpha_2=\alpha a$, $\alpha_3=\alpha b$,
and $\alpha_4=-\alpha a b$.  Second, letting
\[ d = \begin{bmatrix} 0\\0\\0\\1\\ \end{bmatrix} \]
be the fourth standard basis (column) vector of $R^4$,
the standard basis vectors are
$e_1=bac$, $e_2=-bc$, $e_3=ac$, and $e_4=c$.

Thus, with these endomorphisms, our quiver simplifies tremendously: we
only need one arrow in each direction ($c$ and $d$, respectively)
together with two loops $a$ and $b$ at the node corresponding to $M$.
There is also a possibility that more loops are required at the $R$
node as we see in section \ref{ss:laufer}.

Let us see how the relations that we already found translate to this language.
The relations in eq.~\eqref{eq:rel2} are all tautologies, as is easily
verified.  The number of operators $\alpha_i e_j$
which need to 
be considered for eq.~\eqref{eq:rel1} is smaller, and consists of the
ten operators 
\begin{align*}
&c d = \alpha_1 e_4, \quad
c a d = \alpha_2 e_4, \quad
c b d = \alpha_3 e_4, \quad
c a^2d = \alpha_2 e_3, 
\\
&c ab d = -\alpha_4e_4, \quad
c b ad = \alpha_3e_3, \quad
c b^2d= - \alpha_3e_2, 
\\
&c ab ad = -\alpha_4e_3, \quad
c ab^2d = \alpha_4e_2, \quad
c ab^2ad=-\alpha_4e_1.
\end{align*}
There are 45 relations which assert that these all commute
(for example: $cd c ad=c ad cd$).

Finally, the relations in eq.~\eqref{eq:rel3} can be written as:
\begin{align*}
0&=c\left(- abd c+ bd c a- ad c b+ d c ab\right)\\
0&=c\left(- abad c +bad c a-aad c b+ad c ab\right)\\
0&=c\left(- bbad c + bbd c a- abd c b+ bd c ab\right)\\
0&=c\left( abbad c - abbd c a+ abad c b- bad c ab\right)\\
\end{align*}
while those in eq.~\eqref{eq:rel4} can be written as
\begin{align*}
0&=\left(-bbd c ab+bd c aba-ad c abb+d c abba\right)d\\
0&=\left( bbd c b-bd c ba+ad c bb-d c abb\right)d\\
0&=\left(-bad c a+bd c aa-ad c ab+d c aba\right)d\\
0&=\left(-bad c +bd c a-ad c b+d c ba\right)d\\
\end{align*}

There is no reason to suppose that the universal flop can be described
in terms of superpotential since it is of dimension $>3$. In order to
proceed further it is easier to specialize to particular cases.

\subsection{The Morrison--Pinkham example}  \label{ss:MP}

The first examples of flops of length $2$ were given by Laufer 
\cite[Example 2.3]{[L]};
a deformation of Laufer's first example was subsequently found
by Morrison and Pinkham
\cite[Example 10]{[P]}.  Later, Reid \cite[Example (5.15)]{pagoda} 
pointed out that these examples could be put in a standard form,
and it was shown in \cite{flopfactorization} that there is a corresponding
matrix factorization.  We give the matrix factorization for the
Morrison--Pinkham example explicitly.

The Morrison--Pinkham example has equation
\begin{equation}
 x^2+y^3+wz^2+w^3y-\lambda wy^2 -\lambda w^4=0,  \label{eq:MP}
\end{equation}
where $\lambda\in\C$ is a parameter. Although this might appear, at first
sight, to be a continuous family of singularities as one varies
$\lambda$, there are actually only two isomorphism classes of
singularities in the sense of differential equivalence
\cite{AGV:sing}. These cases are $\lambda=0$ and $\lambda\neq0$.

This maps to the
universal flop of length two given in (\ref{eq:univ2}) via
\begin{equation}
\begin{split}
t&=-w\\
u&=y-\lambda w\\
v&=0
\end{split} \label{eq:MPrel}
\end{equation}
giving the matrix factorization
\begin{equation}
\Psi:=\begin{pmatrix}
x&y&z&-w\\
-(y-\lambda w)y&x&(y-\lambda w)w&z\\
-wz&-w^2&x&-y\\
(y-\lambda w)w^2&-wz&(y-\lambda w)y&x
\end{pmatrix}
\end{equation}
\begin{equation}
\Phi:=\begin{pmatrix}
x&-y&-z&w\\
(y-\lambda w)y&x&-(y-\lambda w)w&-z\\
wz&w^2&x&y\\
-(y-\lambda w)w^2&wz&-(y-\lambda w)y&x
\end{pmatrix} .
\end{equation}
In this case, the matrix entries of $\Phi$ generate the ring
\[R=\mathbb{C}[w,x,y,z]/(x^2+y^2+wz^2+w^3y-\lambda wy^2 -\lambda w^4),\]
so we don't need any additional loops at the node $R$.

All of the relations from the previous section hold.  In addition, we get
some new ones, derived from our new relations (\ref{eq:MPrel}).
But perhaps it is easiest just to directly write down the relations among
loops at $R$ in this case.
%
The former ten operators are now reduced to four:
\[
c d = \alpha_1 e_4, \quad
c a d = \alpha_2 e_4, \quad
c b d = \alpha_3 e_4, \quad
c b ad = \alpha_3e_3
\]
and these must pairwise commute:
\begin{align*}
 cd c ad&=c ad cd\\
 cd c bd&=c bd cd\\
 cd c bad&=c bad cd\\
 c ad c bd&=c bd c ad\\
 c ad c bad&=c bad c ad\\
 c bd c bad&=c bad c bd\\
\end{align*}

The remaining six operators now give relations (from $vt$, $uy+2vz$, $ut$, $wz$, $wt$, $uwt$,
respectively):
\begin{equation}
\begin{split}
0&= c \left( ab+ba\right) d\\
c abad&=c\left( bd c b-\lambda
  d c b\right)d\\
c a^2d &= c\left( -bd c + \lambda d c\right)d\\
c ab^2d &= c\left(-ad c \right)d\\
c b^2d &= c\left( -d c \right)d\\
-c ab^2ad &= c\left( -bd cd c
+\lambda d cd c \right)d
\end{split}
\end{equation}

Explicit computation using the matrices directly shows that {\em
  all\/} these relations reduce to
\begin{equation}
\begin{split}
  (b^2 + d c)d&=0\\
  c(b^2 + d c) &= 0\\
  ab + ba &= 0\\
  a^2 + bd c + d c b + \lambda b^2 + b^3 &=
  0.
\end{split} \label{eq:MPmrel}
\end{equation}

It is now a simple observation that our master relations
(\ref{eq:MPmrel}) are obtained as derivatives of the
superpotential\footnote{Some aspects of this superpotential were also derived
  in \cite{Tod2:C-31} using the methods of \cite{Aspinwall:2004bs}.}
\begin{equation}
  \mathcal{W} = b^2d c + \ff12d cd c 
   + a^2b + \ff13\lambda b^3 + \ff14b^4.  \label{eq:WMP}
\end{equation}

Note that the endomorphisms at $M$ comprising elements of $R$
are accounted for by
\begin{equation}
\begin{split}
  w &= -b^2\\
  x &= d c ba + bad c + b^3a\\
  y &= -\lambda b^2 - a^2\\
  z &= -d c a - ad c - ab^2
\end{split}
\end{equation}
It is a then a simple matter to use Macaulay 2 to check that we have
accounted for all possible endomorphisms of $R\oplus M$ by explicitly
computing $\End(R\oplus M)$ as an $R$-module.

We now have the following:
\begin{theorem}
  For the relations given by (\ref{eq:MPmrel}), the quiver associated
  to the noncommutative resolution of (\ref{eq:MP}) is given by
  
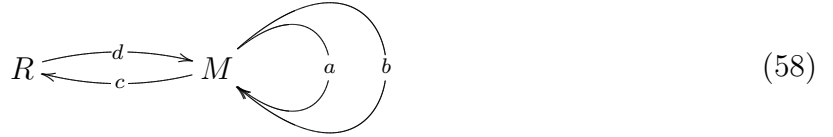
\vspace{10mm}
\begin{equation}
\xymatrix@C=20mm{
  R\ar@/^/[r]|{d\,}&M\ar@/^/[l]|{\,c\,}
          \ar@`{c+(20,20),c-(-0,40)}[]|{\vphantom{E^W_W}a}
          \ar@`{c+(30,30),c-(-0,60)}[]|{\vphantom{E^W_W}b}
}\\[8mm]\label{eq:quivMP}\end{equation}
and we have an associated superpotential (\ref{eq:WMP}).
\end{theorem}

Now that we have a candidate quiver and superpotential we need to
check we have completed the noncommutative resolution, that is, the
global dimension is finite.

Recall that $\mathsf{P}$ is the path algebra of this quiver without
relations imposed and $I$ is the ideal of relations. We need to impose
an admissible order on the monomials of $\mathsf{P}$ and then find a
Gr\"obner basis for $I$. Picking an order at random tends to produce
an infinite Gr\"obner basis.

Let $\C(\lambda)$ be the field of rational functions in $\lambda$. We
may view the path algebra as an algebra over this field.  The ideal of
relations is then quasi-homogeneous if we assign weight 2 to $b$,
$c$, $d$, and $\lambda$; and weight 3 to $a$.  This allows
us to form a weight-lexicographic order on $\mathsf{P}$ by using
weight and then arrow order $a>b>c>d$. The result is that
we have a finite Gr\"obner basis for the relations of the form:%
\footnote{We thank Ed Green for a useful correspondence on this. We
  also used the computer package ``bergman'', developed by 
 J\"orgen Backelin, for computations for the
  Gr\"obner basis.}
\begin{gather*}
ab + ba\\
c b^2 + cd c\\
b^2d + d cd\\
a^2 + b^3 + \lambda b^2+ bd c + d c b \\
ad cd + b^2ad\\
ad c b - b ad c +\lambda b^2a-
  2b^3a -bd c a
  - d c ba\\
c bd cd - cd c b d.
\end{gather*}

Let $L_R$ denote the one-dimensional representation of $A$. Using
(\ref{eq:Lres})  we
have a candidate resolution:
\begin{equation}
\xymatrix@1@M=2mm@C=12mm{
R\ar[r]^{d}
& M \ar[r]^{b^2+d c} & M
\ar[r]^{c}
&R\ar[r]&L_R} \label{eq:LRres}
\end{equation}
\begin{theorem}
  The complex (\ref{eq:LRres}) is exact.
\end{theorem}

We already know this complex is exact except possibly at the first two
terms from section \ref{ss:path}. To check exactness at the first term
we need to show that left-multiplication by $d$ is
injective. Given our choice of term ordering, with $d$ the
smallest arrow, any nontrivial element of the ideal of relations of
the form $d x$ would be detected by a Gr\"obner basis element
left-divisible by $d$. There is no such element and so the complex
is exact at the first term.

Exactness at the second term can be argued as follows. Suppose, in
$\mathsf{P}$, $z=(b^2+d c)x$ is in the ideal $I$. Thus, the
leading term $b^2x$ must be divisible by a leading term of
$\mathcal{G}$. Without loss of generality we may assume $x$ has
already been reduced with respect to $\mathcal{G}$. Inspection of
$\mathcal{G}$ shows that the only possibilities are $\LT(x)=bd
y$ or $\LT(x)=d y$.

Suppose $\LT(x)=bd y$. Then $z$ may be reduced by subtracting
$b(b^2d + d cd)y$. The problem now is that
$bd cd y$ cannot be canceled by anything remaining in
$(b^2+d c)x$ and it cannot be reduced as a leading term.
So we cannot reduce $z$ to lie in $I$.

So it must be that $\LT(x)=d y$. Then $z$ is reduced by subtracting
$(b^2+d c)d y$ to leave $z'$. Now repeat this process
until $z$ is reduced to 0. Since the leading term of $x$ at each stage
was of the form $d y$ we require $x$ to be in the image of
left-multiplication by $d$. Thus (\ref{eq:LRres}) is exact.

Alternatively one may prove exactness at the second term as
follows. Use the Gr\"obner basis above to compute the Hilbert
functions for $R$ and $M$ by enumerating all possible paths. 
One can show that
\begin{equation}
\begin{split}
  H_R(q) &= \frac{1+q^5}{(1-q^2)(1-q^4)(1-q^7)}\\
  H_M(q) &= \frac{1+q^3}{(1-q^2)^2(1-q^7)}.
\end{split}
\end{equation}
We do not include the details here as they are lengthy. One can then
use this to show that the complex (\ref{eq:LRres}) is exact given that
it is exact at all terms except one.  \QED

If $L_M$ is the one-dimensional representation associated to the $M$
node we have a candidate complex quasi-isomorphic to $L_M$:
\begin{equation}
\xymatrix@1@M=2mm@C=15mm{
M\ar[r]_-{f_2}^-{\left(\begin{smallmatrix}c\\a\\b\end{smallmatrix}\right)}
& R\oplus M^{\oplus 2} \ar[rr]_{f_1}^{\left(\begin{smallmatrix}
cd&0&c b\\
0&b&a\\
bd&a&d c+\lambda b+b^2
\end{smallmatrix}\right)} && 
R\oplus M^{\oplus 2} 
\ar[r]_-{f_0}^-{\left(\begin{smallmatrix}d&a&b\end{smallmatrix}\right)}
&M}  \label{eq:resM}
\end{equation}
\begin{theorem}
  The complex (\ref{eq:resM}) is exact.
\end{theorem}

We need to prove $\ker(f_1)\subset\im(f_2)$. The fact
$\ker(f_0)\subset\im(f_1)$ proves that if
\begin{equation}
  ax + by = 0,
\end{equation}
then $x=bz$ and $y=az$ for some path $z$. Suppose
$\left(\begin{smallmatrix}a\\b\\c\end{smallmatrix}\right)$ is in
$\ker(f_1)$. It follows that $b=az$ and $c=bz$ for some path
$z$. This, in turn implies
\begin{equation}
  bd(a-c z)=0.
\end{equation}
We already know that left-multiplication by $d$ is injective. To
prove that left-multiplication by $b$ is injective we need a new
ordering where the arrow $b$ is less than all other arrows.  All
possibilities of the remaining ordering leaves an infinite Gr\"obner
basis. For example if we order $b<d<c<a$ then all
elements of $\mathcal{G}$ of weight $\geq10$ can be shown to be of the
form
\begin{equation}
  c b^{2n}d c + c b^{2n+2}.
\end{equation}
While this cannot be directly exhibited by computer, it is easy enough to apply the
algorithm in \cite{MR1714602} by hand to verify this. Again, no
element of $\mathcal{G}$ is left-divisible by $b$. It follows that
$a=c z$ and we prove exactness of (\ref{eq:resM}) at the second
term.

Since we have shown above that left-multiplication by $b$ is injective
we have that (\ref{eq:resM}) is exact at the first term.\QED

Finally let us recall that we only expect there to be two inequivalent
possibilities; namely $\lambda=0$ or $\lambda\neq0$. Also recall that
the superpotential is only defined up to
$A_\infty$-isomorphisms. These $A_\infty$-isomorphisms correspond to
nonlinear reparametrizations of $a,b,c,d$ as discussed in
\cite{Aspinwall:2004bs}. Suppose we have
a conformal field theory associated with the AdS/CFT
correspondence for this singularity. As such, the superpotential
should be quasi-homogeneous. All this implies that if $\lambda\neq 0$,
the term $\ff14b^4$ in the superpotential is {\em irrelevant\/} and
may be removed by an $A_\infty$-isomorphism. Thus we expect only two
inequivalent superpotentials:
\begin{equation}
  \mathcal{W} = \begin{cases} b^2d c + \ff12d cd c 
   + a^2b + \ff14b^4\quad\hbox{for $\lambda=0$}\\ 
   b^2d c + \ff12d cd c 
   + a^2b + \ff13 b^3\quad\hbox{for $\lambda\neq0$.}\end{cases}
\end{equation}

\subsection{$A_\infty$ Structure}  \label{ss:Ainf}

Let $\DC(\textrm{fdmod-$A$})$ be the full subcategory of
$\DC(\textrm{mod-$A$})$ given by finite-dimensional quiver
representations.  $\DC(\textrm{fdmod-$A$})$ has an
$A_\infty$-structure coming from projective resolutions of vertex
simple objects $L_i$ as was discussed in
\cite{AF:superq,MR2450725}. (We refer to
\cite{ginzburg06:CYa,Lazaroiu:2005da} for an more in-depth discussion
of $A_\infty$-algebras and quivers.) The $A_\infty$-structure encodes the
superpotential. There has been some progress towards equating the
superpotential arising from the quiver relations and the
superpotential arising from this $A_\infty$ algebra \cite{Carqueville:2009xu},
but to date a complete proof seems elusive.

It is a simple matter to verify that the latter
superpotential coincides with that calculated above. Let us put
$\lambda=0$ to simplify the discussion.

Each arrow in the quiver from node $i$ to node $j$ is associated
to a basis element of $\Ext^1(L_i,L_j)$. We represent these arrows
as maps between the projective resolutions. For example, $a$ is given by
\begin{equation}
\xymatrix@M=2mm@C=15mm@R=20mm{
&M\ar[r]_-{f_2}^-{\left(\begin{smallmatrix}c\\a\\b\end{smallmatrix}\right)}
\ar[d]^-{\left(\begin{smallmatrix}0\\1\\0\end{smallmatrix}\right)}
& R\oplus M^{\oplus 2} \ar[r]_{f_1}
\ar[d]^-{\left(\begin{smallmatrix}0&0&0\\0&0&1\\0&1&0\end{smallmatrix}\right)}
 &
R\oplus M^{\oplus 2} 
\ar[r]_-{f_0}^-{\left(\begin{smallmatrix}d&a&b\end{smallmatrix}\right)}
\ar[d]^-{\left(\begin{smallmatrix}0&1&0\end{smallmatrix}\right)}
&M \\
M\ar[r]_-{f_2}^-{\left(\begin{smallmatrix}c\\a\\b\end{smallmatrix}\right)}
& R\oplus M^{\oplus 2} \ar[r]_{f_1} &
R\oplus M^{\oplus 2} 
\ar[r]_-{f_0}^-{\left(\begin{smallmatrix}d&a&b\end{smallmatrix}\right)}
&M
}
\end{equation}
while $b$ is given by

\begin{equation}
\xymatrix@M=2mm@C=15mm@R=20mm{
&M\ar[r]_-{f_2}^-{\left(\begin{smallmatrix}c\\a\\b\end{smallmatrix}\right)}
\ar[d]^-{\left(\begin{smallmatrix}0\\0\\1\end{smallmatrix}\right)}
& R\oplus M^{\oplus 2} \ar[r]_{f_1}
\ar[d]^-{\left(\begin{smallmatrix}0&0&c\\0&1&0\\d&0&b\end{smallmatrix}\right)}
 &
R\oplus M^{\oplus 2} 
\ar[r]_-{f_0}^-{\left(\begin{smallmatrix}d&a&b\end{smallmatrix}\right)}
\ar[d]^-{\left(\begin{smallmatrix}0&0&1\end{smallmatrix}\right)}
&M \\
M\ar[r]_-{f_2}^-{\left(\begin{smallmatrix}c\\a\\b\end{smallmatrix}\right)}
& R\oplus M^{\oplus 2} \ar[r]_{f_1} &
R\oplus M^{\oplus 2} 
\ar[r]_-{f_0}^-{\left(\begin{smallmatrix}d&a&b\end{smallmatrix}\right)}
&M
}
\end{equation}
We now form a differential graded algebra with basis given by the
generators of $\Ext^1(L_i,L_j)$ as above. The product structure is
given by composing maps while one may define a differential
$\mathsf{d}$ in the usual way on chain maps. For example, the
composition $b\cdot b$ is a chain map yielding an element of
$\Ext^2(L_M,L_M)$. This element is zero since it is associated to chain homotopy
given by a map we call $\gamma$:
\begin{equation}
\xymatrix@M=2mm@C=15mm@R=20mm{
&M\ar[r]_-{f_2}^-{\left(\begin{smallmatrix}c\\a\\b\end{smallmatrix}\right)}
\ar[d]^-{\left(\begin{smallmatrix}0\\0\\0\end{smallmatrix}\right)}
& R\oplus M^{\oplus 2} \ar[r]_{f_1}
\ar[d]^-{\left(\begin{smallmatrix}1&0&0\\0&0&0\\0&0&1\end{smallmatrix}\right)}
 &
R\oplus M^{\oplus 2} 
\ar[r]_-{f_0}^-{\left(\begin{smallmatrix}d&a&b\end{smallmatrix}\right)}
\ar[d]^-{\left(\begin{smallmatrix}0&0&0\end{smallmatrix}\right)}
&M \\
M\ar[r]_-{f_2}^-{\left(\begin{smallmatrix}c\\a\\b\end{smallmatrix}\right)}
& R\oplus M^{\oplus 2} \ar[r]_{f_1} &
R\oplus M^{\oplus 2} 
\ar[r]_-{f_0}^-{\left(\begin{smallmatrix}d&a&b\end{smallmatrix}\right)}
&M
}
\end{equation}
Thus we say that $b\cdot b=\mathsf{d}\gamma$.

One can now follow exactly the procedure in \cite{Aspinwall:2004bs} to
extract an $A_\infty$-algebra from this dga and thus find a
superpotential. The result is that one recovers (\ref{eq:WMP}) and
so the two definitions of superpotential agree.

\subsection{Laufer's examples}  \label{ss:laufer}

Now consider the examples described in \cite{[L]}.
Start with an integer $n>0$, and the 
equation\footnote{Laufer presents this with different
variables.  In these variables, the equation reads:
$v_4^2+v_2^3-v_1v_3^2-v_1^{2n+1}v_2=0$ (correcting a typo from \cite{[L]}).}
\begin{equation}
x^2 + y^3 + wz^2 + w^{2n+1} y = 0.  \label{eq:Lauf}
\end{equation}
We can map this to the universal flop of length two via:
\begin{align*}
t&=w^n\\
u&=y\\
v&=0
\end{align*}
Then the matrix factorization is
\begin{equation}
\Psi:=\begin{pmatrix}
x&y&z&w^n\\
-y^2&x&-yw^n&z\\
-wz&w^{n+1}&x&-y\\
-yw^{n+1}&-wz&y^2&x
\end{pmatrix}
\end{equation}
\begin{equation}
\Phi:=\begin{pmatrix}
x&-y&-z&-w^n\\
y^2&x&yw^n&-z\\
wz&-w^{n+1}&x&y\\
yw^{n+1}&wz&-y^2&x
\end{pmatrix}
\end{equation}

The maps in the quiver are:
\begin{equation}
\begin{split}
a&=\begin{pmatrix}0&1&0&0\\-y&0&0&0\\0&0&0&1\\0&0&-y&0\end{pmatrix}\\[3mm]
b&=\begin{pmatrix}0&0&1&0\\0&0&0&-1\\-w&0&0&0\\0&w&0&0\end{pmatrix}\\[3mm]
c&=\begin{pmatrix}x&-y&-z&-w^n\end{pmatrix}\\[3mm]
d&=\begin{pmatrix}0\\0\\0\\1\end{pmatrix}
\end{split}
\end{equation}

The endomorphisms in $R$ at $M$ correspond to
\begin{equation}
\begin{split}
  w &= -b^2\\
  x &= d c ba + bad c + (-1)^m b^{2n+1}a\\
  y &= - a^2\\
  z &= -d c a - ad c - (-1)^mab^{2n}
\end{split}
\end{equation}

An extra loop is required at
$R$ if $n>1$. Let's call it $w$ as it amounts to multiplication by $w$. 
This gives a quiver:
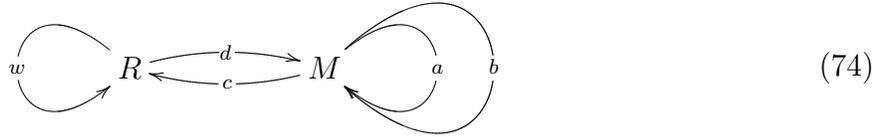
\vspace{10mm}
\begin{equation}
\xymatrix@C=20mm{
  R\ar@/^/[r]|{d\,}
          \ar@`{c+(-20,20),c-(0,40)}[]|{\vphantom{E_0^0}w}
&M\ar@/^/[l]|{\,c\,}
          \ar@`{c+(20,20),c-(0,40)}[]|{\vphantom{E^W_W}a}
          \ar@`{c+(30,30),c-(0,60)}[]|{\vphantom{E^W_W}b}
}\end{equation}
\vspace{10mm}

The superpotential is then
\begin{equation}
  \mathcal{W} = d wc +d c b^2 + a^2b +
\frac{w^{n+1}}{n+1} + \frac{(-1)^n}{2n+2}b^{2n+2}.
\end{equation}

Confirmation that this noncommutative resolution is complete is
similar to section \ref{ss:MP}. Note that the superpotential is
quasi-homogeneous which facilitates computations of Gr\"obner bases, etc.

In the case that $n=1$ we do not require the loop associated to $w$
and the quiver becomes that of (\ref{eq:quivMP}). The superpotential
becomes
\begin{equation}
  \mathcal{W} = d c b^2 -\ff12cdcd + a^2b -\ff14b^4,  \label{eq:L1}
\end{equation}
which is equivalent to the Morrison--Pinkham example with $\lambda=0$
as expected.

\subsection{ A Resolution} \label{ss:frac}

So far we have made no reference to any resolution of the singularities
in question. In each case there are two inequivalent resolutions
related by $x\mapsto -x$. These two resolutions lead to a {\em flop\/}
between the two possibilities.

A resolution can be described purely in terms of a matrix
factorization itself as described in \cite{flopfactorization}. In
general, suppose we have a hypersurface $X\subset\C^N$ described by
$f=0$ with isolated singularity at the origin. Suppose further we have
an $n\times n$ matrix factorization $\Psi\Phi=f$. Away from the
origin, on $X_{\textrm{smooth}}$, the rank of $\Phi$ is $r$. In all
the cases studied in this paper $n$ is even and $r=n/2$.
The kernel of $\Phi$ defines a point in the Grassmannian
$\Gr(r,n)$. The Grassmannian resolution, $\widetilde X$, of $f=0$ is
then defined as the {\em closure} of the point-set
\begin{equation}
  \{(x,v) | x\in X_{\textrm{smooth}},v\in\ker\Phi_x\} \,\subset\,
  \C^N\times \Gr(r,n).
\end{equation}
The space $\Gr(r,n)$ (assuming $n=2r$) is of dimension $r^2$ and is
covered by $\binom{n}{r}$ affine charts using the Pl\"ucker
coordinates. For example, suppose $n=4$. In chart $U_0$ we could
consider matrices
\begin{equation}
  J_0=\begin{pmatrix}1&0\\0&1\\\alpha_{11}&\alpha_{12}\\\alpha_{21}&\alpha_{22}
\end{pmatrix}
\end{equation}
where $\Phi J_0=0$ and in chart $U_1$ we could consider matrices
\begin{equation}
  J_1=\begin{pmatrix}1&0\\\beta_{11}&\beta_{12}\\0&1\\\beta_{21}&\beta_{22}
\end{pmatrix} \label{eq:J1}
\end{equation}
where $\Phi J_1=0$.
The transition functions between these charts are given by
\begin{equation}
  J_0\begin{pmatrix}1&0\\\beta_{11}&\beta_{12}\end{pmatrix} = J_1.
\end{equation}

Some lengthy algebra, described in
\cite{flopfactorization}\footnote{Note that \cite{flopfactorization}
  describes the blowup in terms of $\ker\Psi$ rather than
  $\ker\Phi$. This effectively just changes the sign of $x$.}
can be used to describe the coordinate charts on the blowup.

Suppose we fix on the Laufer examples from section \ref{ss:laufer}.
We will recover the exact description of the charts used by Laufer. To
facilitate this comparison we will need to switch coordinates to a new
set we denote by hats. In chart $U_1$ one gets the equations (from
equations labeled by $\mu_3$, $\mu_{1,2}$, and $\mu_{3,2}$
respectively in \cite{flopfactorization}):
\begin{equation}
\begin{split}
\beta_{2,2}^2+y+\beta_{1,2}^2w&=0\\
\beta_{1,2}y+z+\beta_{2,2}w^{n}&=0\\
\beta_{1,2}w^{n+1}-x-\beta_{2,2}y&=0
\end{split}
\end{equation}
These can be solved:
\begin{equation}
\begin{split}
y&=-\beta_{2,2}^2-\beta_{1,2}^2w\\
z&=\beta_{1,2}\beta_{2,2}^2+\beta_{1,2}^3w-\beta_{2,2}w^{n}\\
x&=\beta_{1,2}w^{n+1}+\beta_{2,2}^3+\beta_{1,2}^2\beta_{2,2}w
\end{split}
\end{equation}
Now relabel the coordinates of $\C^4$ by $w=-v_1$,
$x=(-1)^{n}v_4$, $y=v_2$ and $z=-v_3$. If we assign
$v_1=\hat z_1$, $\beta_{2,2}=(-1)^{n+1}\hat z_2$ and $\beta_{1,2}=\hat w$, then these
solutions become
\begin{equation}
\begin{split}
v_1 &= \hat z_1 \\
v_2 &= -\hat z_2^2 +\hat w^2\hat z_1 \\
v_3 &= -\hat w\hat z_2^2+\hat w^3\hat z_1-\hat z_1^n\hat z_2\\
v_4&= -\hat w\hat z_1^{n+1}-\hat z_2^3+\hat w^2\hat z_1\hat z_2,
\end{split}
\end{equation}
where $(\hat w,\hat z_1,\hat z_2)$ are good affine coordinates for the
resolution of $X$ in the patch $U_1$, and $(v_1,v_2,
v_3,v_4)$ are coordinates of the embedding $X\subset\C^4$. Note that
$v_4^2+v_2^3 -v_1v_3^2 - v_1^{2n+1}v_2=0$
which is the equation for $X$.

To go to patch $U_0$ we define new coordinates $\hat x=\alpha_{12}$
and $\hat y_2=(-1)^{n+1}\alpha_{22}$. We need a third good coordinate
$\hat y_1$ which is a little awkward to spot. The trick is to note that
we need to correctly parametrize the branch of $X$ where $v_1=
v_2=v_4=0$. To this end put $v_3=\hat y_1+\xi$, where $\xi$ is
a function of $(\hat x,\hat y_1, \hat y_2)$ which vanishes when 
$\hat x=\hat y_2=0$. A little algebra yields a solution
\begin{equation}
\begin{split}
  \hat z_1 &= \hat x^3\hat y_1 + \hat y_2^2 + \hat x^2\hat
  y_2^{2n+1}\\
  \hat z_2 &= \hat x^{-1}\hat y_2\\
  \hat w &= \hat x^{-1}
\end{split}
\end{equation}
which is exactly Laufer's set of transition functions. Note that we
manifestly see that the normal bundle of the exceptional $\P^1$ is of
type $(-3,1)$. Note also that $\widetilde X$ is {\em not\/} the total
space of this bundle.

\subsection{Geometrical Interpretation of Branes} \label{ss:geom}

Now that we have the explicit from of the resolution, it is
interesting to see the geometric interpretation of some basic objects
in the D-brane category.

Our tilting objects are $R$ and $M$. Once we have an interpretation of
these in terms of the derived category of coherent sheaves then we can
build any other object we desire. 

$R$ is obvious enough --- it gives the structure sheaf $\O_{\widetilde
  X}$ on the resolution.
$M$ also has a clear interpretation given the explicit form of the
Grassmannian blow-up as we now explain.

For definiteness let us focus on the Laufer example where $n=1$ and
look at the affine coordinate patch $U_1$ with coordinates $(\hat
w,\hat z_1,\hat z_2)$. The matrix $J_1$ given in (\ref{eq:J1}) is
\begin{equation}
\begin{pmatrix}1&0\\-\hat z_2&\hat w\\0&1\\-\hat w\hat z_1&\hat z_2\end{pmatrix}
\label{eq:J1a}
\end{equation}
and has image equal to $\ker(\Phi)=\Im(\Psi)$. Thus, $M$ which is
defined as the cokernel of $\Psi$ can therefore be viewed in the patch
$U_1$ as the cokernel of the matrix $J_1$. Viewed thus as an
$R$-module we may find the coherent sheaf $\cM$ associated to $M$.
The matrix (\ref{eq:J1a}) has constant rank 2 and so $\cM$ is a
locally free sheaf of rank 2 in $U_1$. Similarly $M$ has constant
rank in patch $U_0$ and so $\cM$ is locally free of rank 2 over the entire
resolution.

So, given any object in the derived category of quiver
representations, we may find the corresponding geometric object in the
derived category of coherent sheaves on the resolution by first
writing a free resolution in terms of the projective objects $R$ and
$M$ and then replacing these objects by $\O_{\widetilde X}$ and $\cM$
respectively.

Of particular interest are the two simplest one-dimensional quiver
representations $L_R$ and $L_M$. In physics terminology these are
often called the ``fractional branes''. We have the projective
resolutions of these objects in (\ref{eq:LRres}) and
(\ref{eq:resM}).\footnote{A few sign changes are required to convert
  between the Morrison--Pinkham form and the Laufer form.}

Thus, for example, the object $L_M$ can be represented as the complex
\begin{equation}
\xymatrix@1@M=2mm@C=15mm{
M\ar[r]_-{f_2}^-{\left(\begin{smallmatrix}c\\a\\b\end{smallmatrix}\right)}
& R\oplus M^{\oplus 2} \ar[rr]_{f_1}^{\left(\begin{smallmatrix}
-cd&0&c b\\
0&b&a\\
bd&a&d c-b^2
\end{smallmatrix}\right)} && 
R\oplus M^{\oplus 2} 
\ar[r]_-{f_0}^-{\left(\begin{smallmatrix}d&a&b\end{smallmatrix}\right)}
&M.}  \label{eq:resML}
\end{equation}
But now this is a complex of modules over a {\em commutative\/} algebra
$R=\C[\hat w,\hat z_1,\hat z_2]$ and we may use computer algebra
packages such as Macaulay 2 to determine the cohomology.

Let us number positions in the complex so that the final $M$ is in
position 0. One can compute that the cohomology groups $H^i$
vanish in (\ref{eq:resML}) for $i=-3,-2,0$ and
\begin{equation}
  H^{-1} = R/(\hat z_1,\hat z_2).
\end{equation}
It is worth emphasizing the distinction here. As a complex of modules
over the noncommutative path algebra $A$, the complex (\ref{eq:resML}) is
exact except in position 0 and there the cohomology is $L_M$. However,
as a complex of modules over $\C[\hat w,\hat z_1,\hat z_2]$ it is
exact everywhere except at position $-1$.

One may perform the same computation in the other coordinate patch
$U_0$ and the result is
\begin{equation}
  H^{-1} = R/(\hat y_1,\hat y_2).
\end{equation}
The fact that the cohomology is concentrated in a single position in
the complex implies that this is equivalent, in the derived category,
to a complex with a single nonzero module in one position. It follows
that the geometrical interpretation of $L_M$ is $\cF[1]$, where $\cF$
is a sheaf supported on the exceptional curve $C$ (where $\hat
z_1=\hat z_2=0$) where it is locally free and rank one. The ``[1]''
denotes a shift left in the derived category. Since a locally free
sheaf of rank one on $C$ must be of the form $\O_C(a)$ for some
$a\in\Z$, we have that $L_M$ corresponds to $\O_C(a)[1]$. We determine
$a$ shortly.

A similar computation for $L_R$ yields
\begin{equation}
  H^0 =\begin{cases} R/(\hat y_1,\hat y_2^2)\quad\hbox{in $U_0$}\\
       R/(\hat z_1,\hat z_2^2)\quad\hbox{in $U_1$}\end{cases}
\end{equation}
with all other cohomologies vanishing. It follows that $L_R$
corresponds to some sheaf $\cL_R$ which is an extension
\begin{equation}
\xymatrix@1{
0\ar[r]&\O_C(b)\ar[r]&\cL_R\ar[r]&\O_C(c)\ar[r]&0,}  \label{eq:LMe}
\end{equation}
for some integers $b,c$.

The integers $a,b,c$ can be determined by using the equivalence
between the derived category of quiver representations and the derived
category of coherent sheaves. For example, for quiver representations
it is easy to show that 
\begin{equation}
\Ext^k_A(R,L_M)=0,\quad\hbox{for all $k$}.
\end{equation}
This translates into the statement that
\begin{equation}
\Ext^{k+1}_{\O_{\widetilde X}}(\O_{\widetilde
  X},\O_C(a))=0,\quad\hbox{for all $k$}, 
\end{equation}
which, in turn, implies $H^k(C,\O_C(a))=0$ for all $k$. This implies $a=-1$.

Similarly we know
\begin{equation}
\Ext^0_A(R,L_R)= \Ext^1_A(L_R,L_M)= \Ext^1_A(L_M,L_R)=\C, \quad\hbox{etc.}
\end{equation}
These conditions imply that $b=-1$ and $c=0$ and that the extension
(\ref{eq:LMe}) is nontrivial. One can compute
\begin{equation}
  \Ext^1_{\O_{\widetilde X}}(\O_C,\O_C(-1))=\C,
\end{equation}
which shows that the nontrivial extension (\ref{eq:LMe}) is unique.
This completes the proof of the following:

\begin{theorem}  \label{th:fracB}
If $A$ is the path algebra of the quiver (\ref{eq:quivMP}) subject to
constraints given by the superpotential (\ref{eq:L1}) then there is an
equivalence of triangulated categories given by the functor
\begin{equation}
\mathsf{T}:\DC(\hbox{\rm mod-$A$})\to\DC(\widetilde X),
\end{equation}
such that
\begin{equation}
\begin{split}
  \mathsf{T}(R) &= \O_{\widetilde X}\\
  \mathsf{T}(M) &= \cM\\
  \mathsf{T}(L_R) &= \cL\\
  \mathsf{T}(L_M) &= \O_C(-1)[1],
\end{split}
\end{equation}
where $\cL$ is a sheaf corresponding to the unique nontrivial extension
\begin{equation}
\xymatrix@1{
0\ar[r]&\O_C(-1)\ar[r]&\cL\ar[r]&\O_C\ar[r]&0.}  \label{eq:Lex}
\end{equation}
\end{theorem}

It is perhaps worth emphasizing the fact that if we restrict the
geometry to $C$, then there is no nontrivial extension of the form
(\ref{eq:Lex}) since $\Ext_{\O_C}^1(\O_C,\O_C(-1))=0$. Thus, $\cL$ cannot
be considered to be a locally free sheaf on $C$ pushed forward into
$\widetilde X$ by the inclusion map. That is, there is no description
of the object $L_R$ in terms of vector bundles on $C$.

Now consider the skyscraper sheaf, $\O_p$ of a point $p\in\widetilde
X$, also known as a 0-brane. Given a tilting collection of locally
sheaves of rank $n_0,n_1,\ldots$, it was shown in \cite{AM:delP} that
$\O_p$ corresponds to a quiver with dimension vector
$(n_0,n_1,\ldots)$. This means that the skyscraper sheaf on
$\widetilde X$ is given by a quiver
representation of the form
\vspace{5mm}
\begin{equation}
\xymatrix@C=20mm{
  1\ar@/^/[r]|{d\,}&2\ar@/^/[l]|{\,c\,}
          \ar@`{c+(20,20),c-(-0,40)}[]|{\vphantom{E_W^W}a}
          \ar@`{c+(30,30),c-(-0,60)}[]|{\vphantom{E_W^W}b}
}\\[5mm]\label{eq:qD0}\end{equation}
where the numbers represent the dimension of the corresponding vector
space. In physics language, this is a $\GU(1)\times\GU(2)$ quiver
gauge theory.

In terms of K-theory classes this implies that
$[\O_x]=[L_R]+2[L_M]$. This is entirely consistent with the
identifications of the fractional branes in theorem \ref{th:fracB}.
The moduli space of stable representations of the form (\ref{eq:qD0})
should yield the complete space $\widetilde X$. We will not attempt to
derive this here.

Similarly one cans see that the quiver representation with dimension
vector $(1,1)$ corresponds to $\O_C$ and accordingly has no
(unobstructed) deformations.


\section{Resolutions versus Landau--Ginzburg Theories}
\label{s:rLG}

In this paper we have utilized matrix factorizations in analyzing
D-branes, i.e., the derived category of coherent sheaves, on the
resolution of a hypersurface singularity. This is not, of course, the
first time that matrix factorizations have appeared in the context of
D-branes. It was shown in \cite{Kapustin:2002bi} that D-branes in a \LG\
theory with 2-dimensional superpotential $W$ are described by a
category whose objects are matrix factorizations of $W$. In this
section we give some speculative comments about the possible relation
between these two appearances of matrix factorizations.

Let $f=0$ be a hypersurface singularity in $\C^n$ and let
$R=\C[x_1,\ldots,x_n]/(f)$ and $X=\Spec R$. The category of D-branes
in a \LG\ theory was argued in \cite{Orlov:LG} to be a quotient
\begin{equation}
  \DC_{\textrm{Sg}}(X) = \frac{\DC(X)}{\mf{Perf}(X)},
\end{equation}
where $\mf{Perf}(X)$ is the subcategory of $\DC(X)$ given by bounded
complexes of locally-free sheaves of finite type. This category
depends only on local information of $X$ ``inside'' the singularity as
explained in \cite{Orlov:LG}. This is entirely consistent with the
physics of a \LG\ theory where the classical vacua are given by the
critical points of $f$ and so we expect all physics to be localized at
these critical points in some sense.

The category $\DC_{\textrm{Sg}}(X)$ is triangulated but the
homological grading is $\Z_2$-valued. This is to be expected since
$\DC_{\textrm{Sg}}(X)$ is not a \CY\ category. By taking an orbifold
of a \LG\ theory one can restore a full $\Z$-grading \cite{howa}
and obtain a \CY\ category but this is not desired here.

A string theory or conformal field theory associated with a
$\sigma$-model with target space $X$ is quite distinct from a \LG\
theory on $\C^n$ with worldsheet superpotential $f$. We propose one
should think of the $\sigma$-model as seeing the space ``outside'' any
singularity and the \LG\ theory as seeing the space
``inside''. Indeed, a singularity in $X$ for the $\sigma$-model is a
bad thing --- the conformal field theory will be singular if we
consider deforming a smooth defining equation for $f$ to the singular
case. 

We propose, therefore, that the procedure of a noncommutative
resolution should be viewed as moving D-branes from ``inside'' the
singularity as seen by the \LG\ model to the ``outside'' as seen by
the $\sigma$-model. Initially, before the resolution, it is only the
MCM module $R$ itself that lives on the outside. Note that $R$ itself,
corresponding to the matrix factorization $f=1.f$ is a trivial D-brane
for the \LG\ theory. Now, to perform the resolution we pass the MCM
modules into the tilting module. This makes the D-branes of
singularity visible to the outside $\sigma$-model. At the same time,
since the singularity is resolved, the \LG\ theory itself becomes
trivial. Thus, all that was inside the singularity has passed to the 
outside.\footnote{Some 
similar ideas have been formulated in \cite{arXiv:0811.2439}.}

Whether any specific physics can be attached to this description
remains to be seen but it at least forms a nice picture of the
relation between matrix factorizations for \LG\ theories and matrix
factorization for resolutions.


\section{Discussion} \label{s:conc}

Our analysis of these length two singularities is not complete. In
particular we have not considered the exact nature of the flop. In the
case of length one conifolds, one can show that the flop is
associated with a symmetry of the quiver obtained by exchanged the two
nodes. Since there is no such symmetry in the length two case,
something must change. It would also be interesting to study the
moduli space of the sky-scraper sheaf quiver (\ref{eq:qD0}).

Another interesting question concerns massless D-branes on the
singularity. On a simple conifold, either of the two vertex-simple
fractional branes, $\O_C$ or $\O_C(-1)[1]$, become massless (depending
on the $B$-field) as $C$ is blown down. What happens in the cases
studied in this paper? Can the length two fractional brane $L_R$
become massless? The usual way to analyze this would be to find some
local mirror picture and find some Picard--Fuchs system, or use
something like the monodromy ring of \cite{AP:decomp}. We do not know
how to do this for the cases at hand. 

Additionally one can consider extremal transitions. The Milnor numbers
for the singularities we consider are quite large. For example, the
Milnor number of the Laufer case (\ref{eq:Lauf}) is given by
$6k+5$. This means the extremal transition is asssociated to one
deformation of K\"ahler form (blowing up $C$) and $6k+5$ deformations
of complex structure. It would be interesting to see if such a local
transition could be see in a compact \CY\ threefold.

Obviously one may also try to apply our methods to higher length
cases. Whether our method provides any particular computational
efficiency over other methods may be a matter of taste. However, it does
provide a nice picture of how D-branes pass between \LG\ theories and
$\sigma$-models when one resolves the singularity.

\section*{Acknowledgments}

We wish to thank R.~Eager, E.~Green, and E.~Miller for invaluable discussions. 
P.S.A.\ thanks the Kavli Institute for Theoretical Physics
and D.R.M.\ thanks the Aspen Center for Physics for hospitality at
various stages of this project.
This work was partially supported by
NSF grants DMS--0606578 and  DMS--0905923.
Any opinions, findings, and conclusions or recommendations expressed in this
material are those of the authors
and do not necessarily reflect the views of the National Science Foundation.


\ifx\undefined\bysame
\newcommand{\bysame}{\leavevmode\hbox to3em{\hrulefill}\,}
\fi

\end{document}